FUEL CELLS

# Proton channels

The structure of Nafion, the polymer electrolyte membrane used in some fuel cells, has been extensively debated over recent decades. Now, a newly proposed model reveals the nanoscale arrangement that could explain the excellent transport properties of the material.

Olivier Diat* and Gérard Gebel
are in the UMR 5819, Structures et Propriétés d'Architectures Moléculaires, CEA Grenoble, 38054 Grenoble cedex 9, France.
*Present address: UMR 5257, ICSM - CEA Marcoule, BP 17171, 30207 Bagnols s/Céze cedex, France.

Transport is a major source of pollution in cities. As a result, the development of cars and buses which generate low or negligible level emissions of greenhouse gases is an environmental priority. Two promising technologies compete for this niche: Batteries and fuel cells. The latter, and the topic of discussion here, are based on scientific principles known for about two centuries, but the technology has emerged only recently.[1,2] Among the few established fuel cell designs,[1,2] those based on a polymer electrolyte membrane (PEM) and di-hydrogen are promising candidates as power sources for cars as they function in the appropriate operating temperature range (80-120°C) (Figure 1). To date, the best performing PEMs in $H_2/O_2$ fuel cells have been composed of Nafion, a perfluoro-sulfonated copolymer developed by DuPont in the 1960s.[3,4] Despite all the attention Nafion has received in the industry, however, the hierarchical structure of the membrane remains a mystery and, as a result, the mechanisms of the reactions involved are not well understood. On page XX of this issue Klaus Schmidt-Rohr and Qiang Chen enter this structural debate and, by the proposal of a new model which supports the analysis of published X-ray scattering curves, show convincing support for the tubular structure of Nafion.[5] This model gives some interesting arguments for solving, where previous ones have failed, the excellent transport properties of hydrated Nafion.

A fuel cell is an electrochemical converter – chemical energy from the fuel, hydrogen, is converted into electrical energy in the presence of atmospheric oxygen. More specifically, a PEM fuel cell operates using the following simple principles (Figure 1): a hydrogen molecule diffuses through a grooved polar plate and a gas diffusion layer (a carbon tissue), and oxidises when it approaches an active layer at the anode to give two protons and two electrons. The protons pass through the PEM and recombine with the electrons in the active layer of the cathode via catalytic reduction of oxygen to produce water and heat. Proton diffusion, and hence the performance of the fuel cell, is very sensitive to the relative hydration of the membrane-electrodes assembly (MEA). It is thus essential to prevent both drying and flooding of the MEA; to control this, 'water management' technology is used.[6]

Nafion has many characteristics which make it suitable for use as a membrane in fuel cells, most importantly, when hydrated it exhibits high intrinsic proton conductivity and it is chemically and electrochemically stable in an acidic and oxidizing environment. Also, it has low permeability to gas reactants and excellent mechanical properties including flexibility, ductibility and water swelling capacity. As a result, it is better than numerous hydrocarbonated ionic polymers that have been proposed as alternative PEMs.[7,8] Despite this superiority and the considerable research undertaken, including analysis using several characterisation techniques (Fig. 2), the hierarchical structure of the membrane and the mechanism of cation exchange remain poorly understood. Numerous studies demonstrate that the Nafion membrane exhibits a bicontinuous nanostructure involving a hydrophobic matrix and ionic domains that swell on hydration.[7] But, as yet, there is no consensus on its mesoscopic structure, dependence on the fabrication process, thermal history and aging properties. This problematic characterisation is primarily a result of the lack of long range structural order in Nafion, and, because it is a soft material, produced in non-equilibrium conditions, it is sensitive to mechanical stress. In spite of this, numerous works have demonstrated a strong relationship between structure and transport properties.[9] However, the definitive length- or time-scale in the proton diffusion process that determines water and ion transport remains unclear.[8]

The two-dimensional model developed by Schmidt-Rohr and Chen captures the essential physical features of Nafion and numerically recovers the shape of X-ray scattering data extracted from the literature. Using some arguments of symmetry, they described a new structure comprising an array of oriented ionic nanochannels embedded within a locally aligned polymer matrix. The strong molecular interactions between the polymeric backbones make the cylindrical walls stiff and stable. Moreover, this matrix is able to swell perpendicular to the bundle direction when water is sorbed. The confined ions can thus diffuse easily in this tubular nanostructure. This model is in agreement with the observed variation of the scattering intensity as a function of the amount of contained water and could allow for the anisotropic optical and mechanical properties.[10,11] In contrast to previous work, this analysis is not a superposition of several scattering contributions; for example, those from ion clusters with diverse shape and distribution, from crystalline and amorphous domains. It takes into account the supramolecular organisation of the polymer chains as a whole. However, as this work is a numerical analysis, the energy minimum for the system is not found, as a result, the finer details of the self-association of the polymer remain unclear. The comb-like structure of the polymer chain is perhaps a necessary and sufficient condition but further studies are required to understand this association process – this information is essential for the design of new polymers.

Nevertheless, if this approach developed by Schmidt-Rohr and Chen is extended to other ionomer materials for comparison and more specifically to model systems,[12] then it will become fruitful. Of course, this requires scattering data, over a large range of scattering angles, which can only be obtained on large instruments.

At present, the widespread industrial development of fuel cells is somewhat impeded by several limitations, in particular the high cost and short lifetime of some components. Schmidt-Rohr and Chen's understanding of the benchmark membrane could aid the future design of ionic polymers and nanocomposites for PEM applications and ensure lower fabrication costs, longer lifetimes and higher operation temperatures. We can also imagine that a specific development of a fabrication process to obtain a macroscopic orientation of these proton channels will make Nafion an ideal PEM material.

**Legends**

Fig. 1: Overview on an open zero-emission car (NISSAN) and a schematic view of a single $H_2$-fuel cell extracted from a stack. This is a layered assembly: the polar plates that are bipolar in a stack, the electrodes including the gas diffuse layers (GDL) and the active layers, and the heart of the cell, the polymer electrolyte membrane (PEM). Under operation, a water concentration gradient within the MEA results from the different water and ions transport mechanisms.

Fig. 2: View of a Nafion membrane at the centimetre scale (a), at the micron scale (b) using atomic force microscopy (AFM) technique, and at the nanometre scale (c) using an artistic representation. A non exhaustive series of probing techniques are listed as a function of the scaling and in two categories: **static** with optical, electron and atomic microscopies (OM, SEM-TEM, AFM respectively) and small and wide angle scattering techniques (SAS and WAS resp.). **dynamic** with the quasi elastic neutron scattering (QENS), various NMR and diffusion techniques.

**fig. 1**

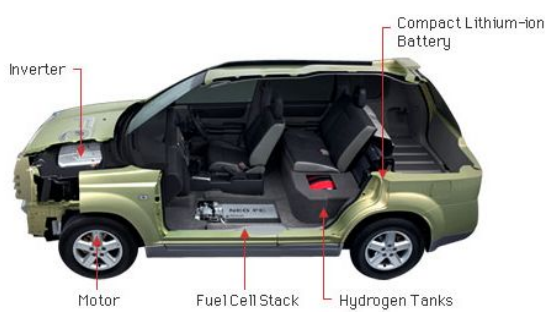

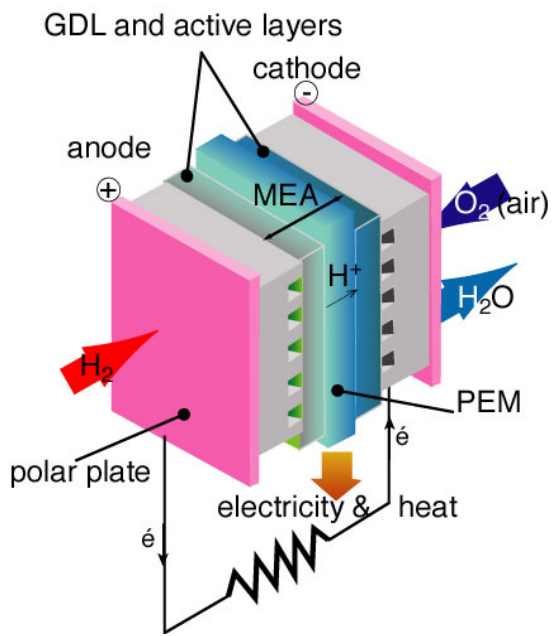

**fig 2 :**

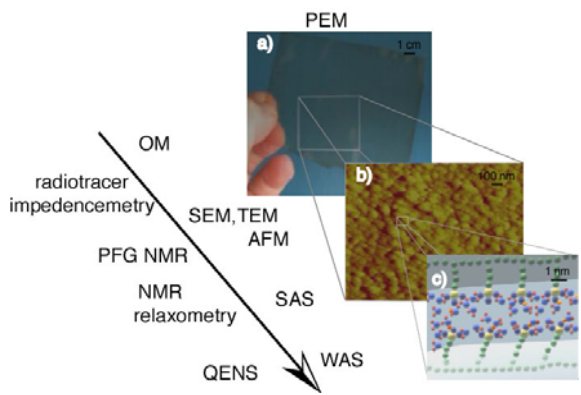